# BioGAN: An unpaired GAN-based image to image translation model for microbiological images

Saber Mirzaee Bafti[a,*], Chee Siang Ang[a], Gianluca Marcelli[a], Md. Moinul Hossain[a], Sadiya Maxamhud[b], Anastasios D. Tsaousis[b]

[a] *School of Engineering and Digital Arts, University of Kent, Canterbury, CT2 7NZ, UK*
[b] *Laboratory of Molecular & Evolutionary Parasitology, RAPID group, School of Biosciences, University of Kent, Canterbury, CT2 7NJ, UK*

**Abstract**

*Background and objective:* A diversified dataset is crucial for training a well-generalized supervised computer vision algorithm. However, in the field of microbiology, generation and annotation of a diverse dataset including field-taken images are time-consuming, costly, and in some cases impossible. Image to image translation frameworks allow us to diversify the dataset by transferring images from one domain to another. However, most existing image translation techniques require a paired dataset (original image and its corresponding image in the target domain), which poses a significant challenge in collecting such datasets. In addition, the application of these image translation frameworks in microbiology] is rarely discussed. In this study, we aim to develop an unpaired GAN-based (Generative Adversarial Network) image to image translation model for microbiological images, and study how it can improve generalization ability of object detection models.

*Methods:* In this paper, we present an unpaired and unsupervised image translation model to translate laboratory-taken microbiological images to field images, building upon the recent advances in GAN networks and *Perceptual* loss function. We propose a novel design for a GAN model, BioGAN, by utilizing *Adversarial* and *Perceptual* loss in order to transform high level features of laboratory-taken images of *Prototheca bovis* into field images, while keeping their spatial features.

*Results:* We studied the contribution of *Adversarial* and *Perceptual* loss in the generation of realistic field images. We used the synthetic field images, generated by BioGAN, to train an object-detection framework, and compared the results with those of an object-detection framework trained with laboratory images; this resulted in up to 68.1% and 75.3% improvement on *F1-score* and *mAP*, respectively. We also present the results of a qualitative evaluation test, performed by experts, of the similarity of BioGAN synthetic images with field images.

*Conclusion:* An unpaired image-to-image translation model for transforming the lab-taken microbiological image to field-like images is proposed. Synthetically generated images by our model are compared with the generated images by standard multi-purpose models. Codes are publicly available at https://github.com/Kahroba2000/BioGAN.

## 1. Introduction

Computerized image processing in biological images has drawn the attention of researchers due to its potential in various applications, including cell counting [1] and disease diagnosis [2,3]. Most state-of-the-art object detection algorithms are built on deep neural networks, which have the challenge of needing a large dataset, often also requiring high quality annotations. Furthermore, when it comes to training object detection algorithms, there are additional requirements, such as having a diverse dataset that covers a wide range of possible conditions (i.e. different brightness, contrast, resolutions, etc.). In microbiology, images collected in the field (i.e. in real growing media including water, stool, etc.) have been used in the past to train object detection algorithms, in order to analyze biological systems in real conditions; these studies have had varying degrees of success [4,5]. Having a diverse dataset including field taken images can lead to a well-generalized object detection model, but as one can imagine, collecting a large, diverse dataset can be expensive, tedious, and in some cases impossible. In the context of the current study, i.e. parasite detection in the field in low income countries, these algorithms need to perform well on images, of vastly varying quality, which are collected using low-cost portable devices (e.g. a smartphone). Although bio scientists are increasingly sharing data online (e.g. any bio data archive), most datasets consist of laboratory-taken data, due to the inherent challenges of collecting field data.

Image-to-image translation (I2IT) refers to techniques that map (transform) an input image, $x$, to a target output image, $y$, ($y = f(x)$) [6]. Such techniques have been widely implemented to tackle challenges in different domains, e.g. translation of aerial images of natural landscape into city-street maps or translation of daytime images into nighttime images, or translation of young faces into aged faces [7–9].

*  Corresponding author. Saber Mirzaee Bafti.
   *E-mail address:* sm2121@kent.ac.uk

Prior to the development of GANs (Generative Adversarial Networks) [10], various traditional techniques based on machine learning approaches have been developed to tackle different challenges including colorization, de-noising, etc. For example, [11] have utilized two different deep convolutional neural networks to extract high-level, mid-level and global features from a grayscale input image and to colorize it. In this approach, the optimizer penalizes the difference between ground truth colorful and input image through backpropagation. However, according to [12] and [13], minimizing Euclidean distance between generated synthetic image and ground truth (GT) in traditional approaches often leads to blurry images. Fortunately, GANs [10] have revolutionized I2IT, because, unlike the traditional approaches which treat pixels independently, GANs aim to find the mapping function between input images and output images, instead of minimizing Euclidean distance. This helps to block low quality images as they will be easily distinguishable by discriminators as fake images.

Following the progress of GAN networks, conditional GANs (cGAN) [14] (i.e. adding the condition as an input to both generator and discriminator) has gained momentum in the field of image translation too. Conditioned generators and discriminators help faster convergence in the training process, and also generate more controllable synthetic images [7]. Conditional GAN networks have tackled various challenges including photorealistic image generation from semantic segmentation [15], domain transfer in fashion image (e.g. Generation or changing the subject's dress in input image) [16], prediction of lost frames in a video stream (i.e. in order to increase framerate) [17], style transferring (e.g. adopting the texture of one image to another) [18], to name a few. Despite the impressive success of these studies, requirements of a paired dataset (i.e. input images and their corresponding output images) to train the model is a deterring barrier for training and utilizing them. To overcome this, the idea of using cycle-consistency loss to train GAN-based image translators with unpaired data has been proposed by [7]. Cycle-consistency loss is made of two forward and backward consistency objectives, which aim to reach $F(G(x)) = x$ in the forward consistency objective, and $G(F(y)) = y$ in the backward consistency objective. The cycle-consistency loss approach has led to several new architectures [8,9,15,19–24]. Despite intensive research of I2IT in general domains, its application in medical and microbiological images is still in its infancy. For instance, Karim et al. [25], developed a cGAN network to minimize the high level features discrepancies between CT (Computed Tomography) and MRI (Magnetic Resonance Imaging) images to highlight more meaningful features for the clinicians. In another study, [26], has developed a new GAN network (known as MI-GAN) to generate synthetic retinal images from random noise vectors in order to enlarge and diversify image datasets. Based on the concept of cycle-consistency loss [7], [27] developed an I2IT framework for synthetically generating Covid-19 diagnosed chest X-ray images. In this way, the challenges of training deep neural networks with little data were overcome. The use of I2IT techniques to deal with lack of data has been applied to other medical imaging problems, such as skin lesion detection [28] as well, however, there are only a few studies that have applied GAN I2IT to microbiological images.

Bailo et al., [29], implemented a novel GAN network for red blood cell image augmentation, in which they have trained two cascaded generators, where the first one generates random instance masks while the second generator translates the instance masks into synthesized blood cells. As another application of I2IT models in microscopy images, [30] proposed an I2IT model based on the idea of Cycle-consistency [7], to artificially staining histological images, or transforming dead phytoplankton cells to living cells [31]. Our finding shows that application of I2IT models in microbiology is still very limited. One such possible application is the translation of laboratory-taken to field-taken images. As shown in Fig. 1, the visual characteristics of microbiological images (of the *Prototheca bovis* parasite) taken in the laboratory are significantly different from field images, due to different reasons, including different photography conditions, image acquisition devices, and most critically the growing media such as water, stool, soil, etc., which can even affect cells' morphology.

Considering the significant difference in cell morphology between laboratory-taken and field-taken microscopic images (i.e. Fig. 1), the effort to develop object (parasite) detection algorithms for microscopic images is hampered by the limited access to field images. So in this paper, we explore whether and how I2IT can be a reasonable solution to generate synthetic field-like images using laboratory-taken images. This is especially challenging because most of the I2IT models require paired data sets which is almost impossible to generate in large numbers for microbiology images. Therefore, we aim to propose a new unpaired GAN-based image-to-image translation design for microbiological image translation, inspired by previous work including [7] and [32]. More specifically, in this study we aim to demonstrate a novel method of using I2IT, which translates laboratory-taken images into synthetic field images to overcome the challenge of accessing field images to train object detection algorithms, which can be useful in parasitic disease detection.

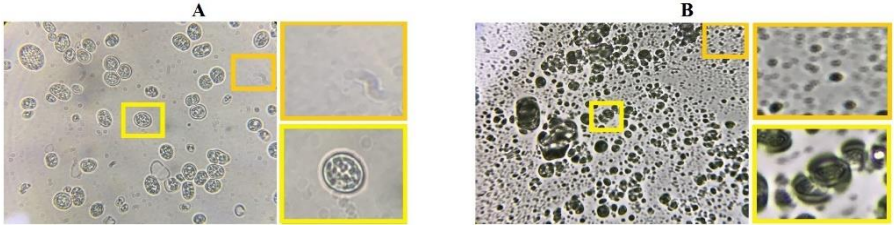

**Fig. 1**. Example images of *Prototheca bovis* that are taken in the laboratory (A) and field environment (B). Moving from laboratory image to field image, both background texture (orange box) and target objects texture (yellow box) change. (Colourful)

## 2. Methods

The backbone of our model is based on a GAN network with a new loss function as shown in Fig. 2. The first two subsections present the architecture of our proposed GAN network (2.1) and the loss functions (2.2). Subsection 2.3 explains the training procedures and the hyper parameters of our proposed model (BioGAN).

*2.1. Model architecture*

I2ITs can be viewed as a function to map an input image to an output image that carries most or parts of spatial features with different appearance. Fortunately, this mapping function is very similar to what has been done in GAN [10]. Thus, inspired by [7,32,33], we utilized a GAN network with a generator and discriminator as discussed as follows. In our proposed approach, a new sort of *Adversarial* plus *Perceptual* loss has been implemented to encourage the generator to learn to create more realistic synthetic images from unpaired data. Furthermore, the backbone of the script including the generator and discriminator has been adopted from [7] and [34].

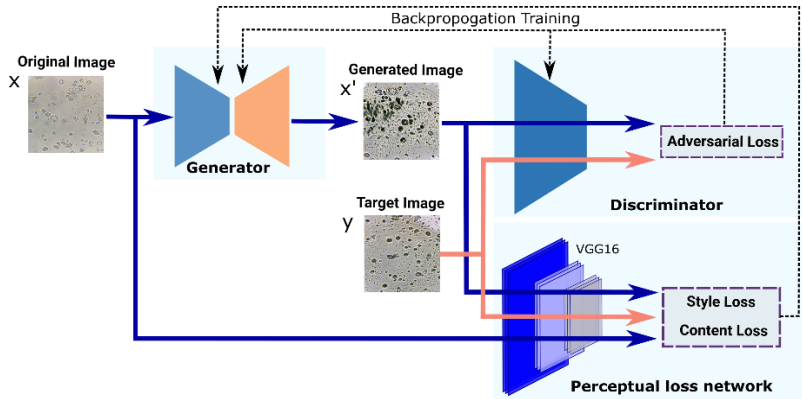

**Fig. 2.** Overview of the proposed model (Colourful)

**Generator**: To map two high-resolution images from one domain to another, a variety of transformers based on neural networks have been developed. Encoder-decoder architectures [35] and residual networks (Res-Net) [36] are two networks widely used in previous studies where the input image should be passed through all layers to reach the end layer. Due to the fact that in I2IT a big portion of the low-level structural features are shared between input and output images, it would be desirable to directly pass these features to the output to avoid any possible distortion. For this purpose, U-Net [37] with skip connections to bypass bottleneck layers is a common technique that has been utilized in studies such as [6] and [25].

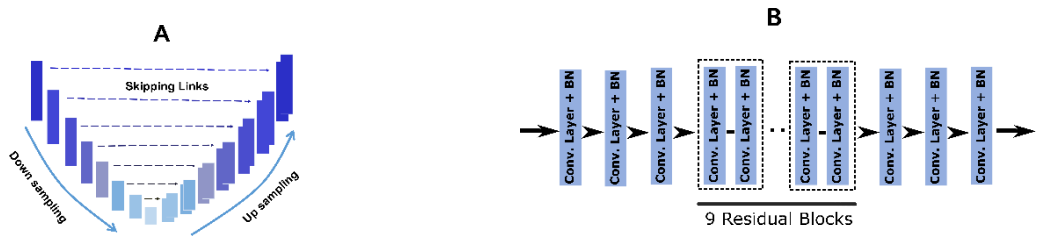

**Fig. 3**. Two different generator architectures: U-Net (A) and Res-Net (B)

We implemented two separate generator architectures based on the Res-Net and U-Net. The implemented Res-Net contains three convolutional layers followed by nine residual blocks, further two convolutional layers and another deconvolution layer (see fig. 3.B). Apart from the last convolution layer, all previous Conv/Deconv layers are followed by a *ReLU* (Rectified Linear Unit) activation function and batch normalization (i.e. the last convolutional layer is followed by a Tanh activation function). For the U-Net, eight down-sampling, and eight up-sampling layers were implemented as shown in Fig. 3.A.

We compared qualitatively Res-Net against U-Nets by visually evaluating the output images. As shown in Fig. 4, as expected, the generated synthetic images with U-Net seem to have resembled more precise and sharper contents' objects in comparison with the Res-Net that has passed the input image through all layers.

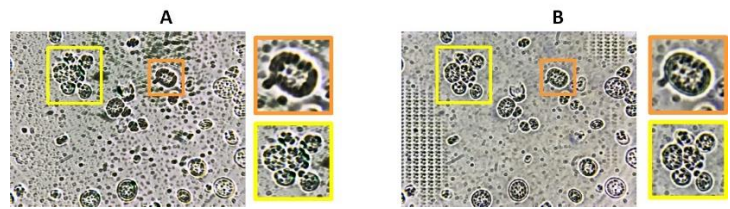

**Fig. 4**. Generated images via Res-Net (A) vs U-Net (B). Due to passing the spatial features through the skipping link in the U-Net, it produces sharper content. (Colourful)

Although the U-Net generator has shown more capability in transferring meaningful spatial features (see yellow boxes in fig. 4), our observation shows that it fails to translate low frequency background which is very common in the laboratory microbiological images (see pixelated background in fig. 4.B). This is due to the implementation of U-Net with the kernel size of 4 and stride of 2 on convolutional layers, leading to a pixellation in low-frequency background regions as depicted in fig. 5.A. To resolve this issue along with keeping the advantage of the U-Net generator over Res-Net, a modified version of the generator with kernel size of 3 and stride of 1 was implemented. Results show that using convoluted kernels with smaller stride gaps leads to a better reconstruction as shown in Fig. 5.B.

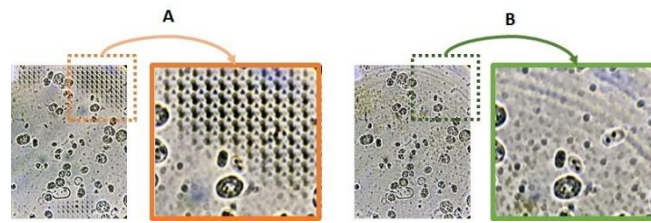

**Fig. 5.** Examples of the U-net generated image with stride of 2 (A) and 1 (B). (Colourful)

**Discriminator**: Discriminator is a binary classifier that is responsible for differentiating between synthetic $x'$, and target image, $y$. Conventional L2 and L1 loss functions in classical discriminators lead to blurry images, and also fail to encourage generating high-frequency crispness [6]. To model the high frequency sections of the input image, [6] suggested focusing on the structure of patches in each image, instead of looking at the entire image as a whole. Unlike full-image discriminators that pass a fix-sized input image through a fully convolutional network, PatchGANs discriminator gets an arbitrary sized input image and encourages the generator to penalize the structure at the patch scale. Because of that, PatchGANs discriminator can be understood as another level of style/texture loss function on top of the style loss function. Thus following [6,7,18,38], we used a $70 \times 70$ PatchGANs to classify synthetic and target (i.e. real field-taken) images.

*2.2. Loss functions*

One of the main objectives in microscopic I2IT is to transfer the texture of the entire input image to the target image, similar to what has been done in [32,33,39,40]. *Perceptual* loss [32,33,39,40] has shown impressive success in encouraging convolutional neural networks to transfer high level features of an input image to others. However, in the context of parasite segmentation, apart from the image texture, lower level features including the appearance of the individual cells in microbiological images are also important. Thus, our model aims to transfer the texture of the field-taken images to laboratory-taken images, along with the appearance of the parasites. The structural features (i.e. cell's outline) of the input image should remain constant. In the following two subsections two elements of our global loss function will be discussed in detail: *Adversarial* loss and *Perceptual* loss.

2.2.1 Adversarial loss

The key loss function of GANs is *Adversarial* loss, which represents the probability of error of an image, i.e. whether it is real or synthetic. However, conditioning *adversarial* loss function with some extra information (for example, the label of the synthetic hand-written digit you want the generator to generate on the MNIST dataset, or the conditions you want the synthetic image to meet) helps to minimize the difference between the synthetic image generated by the generator and the target image [6] as follows:

$$L_{adv}(G,D) = \mathbb{E}_{x,y}[Log D_{(x,y)}] + \mathbb{E}_{x,z}[Log(1 - D(x, G_{(x,z)}))] \qquad (1)$$

, where *x, y* are input and target images, *z* represents the conditional variant. The training objective is:

$$\min_{G} \max_{D} L_{adv}(G,D)$$

Unlike classical I2IT models which aim to penalize the Euclidean distance between the input and target images' pixels, conditional adversarial loss looks at the similarity of two images from a higher level as a whole. However, for faster convergence, and for encouraging the model to generate a less blurry image, a combination of the adversarial loss with traditional loss function (including L2 or L1) [6,12], can be used, although paired data is required for this objective. Therefore, for faster convergence, and to avoid blurry/low contrast images, we have utilized a pre-trained *VGG16* network to create style reconstruction and content reconstruction loss [32,33,39] as explained in the following subsection (2.2.2).

2.2.2 Perceptual loss

*Perceptual* loss is another element included into *BioGAN* loss function. *Perceptual* loss is a non-adversarial loss function, introduced by [32,33,39,40], that measures *Perceptual* differences between the two images' content and style. A perceptual loss function computes the sum of all the squared errors between all pixels within an image, unlike a pixel-wise loss, which computes the sum of absolute errors between pixels. In other words, the perceptual loss function can be used to analyze two images from a higher level, for example, examining the difference in the textures of the images. It is therefore possible to use *Perceptual* loss to transmit high-level features of images to others while maintaining their spatial features, including their main contents.
In *BioGAN* due to the absence of paired data, encouraging the generator to generate a decent un-blurred output without pixel-wise optimization is challenging. In our work, in order to compensate for the absence of loss in pixel level and for faster convergence of the generative G's loss, we use *Perceptual* loss [32,33,39,40] to map the style of the target image and the content of the input image. *Perceptual* loss looks at the discrepancy of style and content of images from a higher level, which is different from pixel-level loss. To transfer meaningful features from the style of the target images $y$ to the input image $x$, the global *Perceptual* loss from the combination of *Style Reconstruction loss* and *Content Reconstruction loss* has been utilized, as they are discussed below.

**Style reconstruction loss:** Comparing laboratory-taken microscopic images with field-taken images, the high-level characteristics (e.g. style/texture) of the images can be significantly different, due to the nature of the media that cells have been grown in. To minimize the style discrepancy between the target image and synthesized image, we use a pre-trained VGG16 image descriptor, containing five convolutional blocks including two to four layers, similarly to [32], which reconstructs the output image's styles with respect to the target image. The style is reconstructed from a combination of different layers of convolutional blocks, including '*Relu1_1*', '*Relu2_1*', '*Relu3_1*', '*Relu4_1*', and '*Relu5_1*' as shown in Fig. 6. Each layer has $N_l$ feature maps (i.e. filters) with size $M_l$, where $M_l$ is the height of the feature map times the width of the feature map. The correlation between the style of the output image and target image is represented by a Gram matric $G_{j,i}^l$ which is the inner product between the vectorised feature maps $i$ and $j$, within layer $l$, as defined below:

$$G_{j,i}^l = \sum_k F_{ik}^l F_{jk}^l \qquad (2)$$

where $F_{ik}^l$ is the activation of the $i^{th}$ filter at position $j$, within layer $l$ [32]. Let $G^l$ and $A^l$ be the output and the target images' style representatives at layer $l$, then the contribution of each layer in the style reconstruction is computed as below:

$$E_l = \frac{1}{4N_l^2 M_l^2} \sum_{i,j} (G_{j,i}^l - A_{j,i}^l)^2 \qquad (3)$$

and the style reconstruction loss is defined as below:

$$L_{style} = \sum_{l=0}^{L} \lambda_{sl} E_l \quad (4)$$

where $L$ is the total number of layers and $\lambda_{sj}$ is the weight of contribution of $l^{th}$ layer in global loss. As shown in Fig. 6, the reconstructed style from high level layers (e.g. starting from *Relu1-1* from Fig. 6) results in smaller-scale structure reconstruction. We have utilized five layers (*'Relu1_1', 'Relu2_1', 'Relu3_1', 'Relu4_1'*, and *'Relu5_1'*) for style reconstruction loss, as they produced the most visually consistent style.

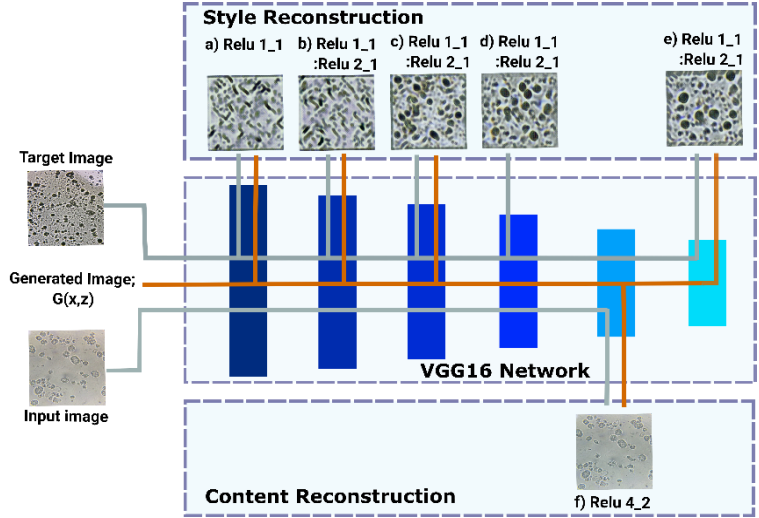

**Fig. 6.** Perceptual loss network to measure two elements: style reconstruction and content reconstruction. Style reconstruction, from different layers of the pre-trained feature extractor model of VGG16, has been done via a) 'Relu1_1' b) 'Relu1_1', 'Relu2_1' c) 'Relu1_1', 'Relu2_1', 'Relu3_1' d) 'Relu1_1', 'Relu2_1', 'Relu3_1', 'Relu4_1' e) 'Relu1_1', 'Relu2_1', 'Relu3_1', 'Relu4_1', and 'Relu5_1' layers. Style reconstruction from higher level conveys larger-scale style structure. Content reconstruction has been done via f) 'Relu4_2'. (Colourful)

**Content reconstruction loss:** Due to the absence of paired images for pixel reconstruction loss (i.e. reconstruction of content at pixel level), and due to the importance of transferring meaningful spatial features from the content of the input image to the output image, *Content loss* aims to minimize the content discrepancies between input and synthetic images. Like the style reconstruction, for content reconstruction, the input image and the output image's content are represented by Gram matrices $G^l$ and $P^l$, respectively. So, the *Content loss* is represented by Eq. (5):

$$L_{content} = \frac{\lambda_{cl}}{2} \sum_{i,j} (G_{j,i}^l - P_{j,i}^l)^2 \quad (5)$$

where $L$ is the total number of layers (i.e. in this case one) and $\lambda_{cl}$ is the weight of the layer contribution. Content reconstruction from lower layers of the feature extractor preserves the main content with original properties, while deeper reconstruction would slightly disturb the high level features of the contents (i.e. colour, shape, texture, etc.) [6,11,41]. In this study, we have reconstructed the content from layer *Relu4-2* as [34].

*2.3. Training*

To train BioGAN, we have used the collected field images as target images, and laboratory-taken images as input ones. BioGAN was trained end-to-end via a min-max optimization task upon the following global loss function:

$$L_{Generator} = \lambda_A L_{Adversarial} + \lambda_S L_{Style} + \lambda_C L_{Content} \quad (6)$$

All lost elements in Eq. 5 are weighted by $\lambda$ in order to tune the influence of each of them on global loss, $L_{Generator}$. The parameters, $\lambda_A$, $\lambda_S$, and $\lambda_C$, are chosen to be $10^4$, 1.0, and 0.4, respectively, after more than 50 trials. Table 1 shows the training pipeline of our model.

**Table 1.** Training pipeline of our model

---
**Require**: *Unpaired training datasets* $\{(x_j, y_j)\}_{j=1}^{T}$
**Require**: *A selected target style image from target images Y*
**Require**: *Training with* $\#_{epoch} = 100$, $\lambda_A = 10e3$, $\lambda_S = 1.0$, $\lambda_C = 0.4$
**Require**: *Pre-trained model of VGG16*

1: **For** n=0, 1, .... $\#_{epoch}$ do:
2:    **For** m=0,1, T do:
3:       **For** k=0, 1, $\#_{iteration}$ do:
4:          $L_{Adversarial} \leftarrow -Log(D(G_{(x_m,z)}, x_m)$
5:          $L_{Style} \leftarrow \sum_{l=0}^{L} \lambda_{sl} \cdot E_l$
6:          $L_{Content} \leftarrow \frac{\lambda_{cl}}{2} \sum_{i,j}(G_{j,i}^l - P_{j,i}^l)^2$
7:          $\theta_G \leftarrow^+ \lambda_A L_{Adversarial} + \lambda_S L_{Style} + \lambda_C L_{Content}$
8:       **End**
9:       $L_{Discriminator} \leftarrow Log(D(G_{(x_m,z)}, y_m)) + Log(1 - D(G_{(x_m,z)}, x_m))$
10:      $\theta_D \leftarrow^+ L_{Discriminator}$
11:   **End**
12: **End**

---

Generally, when an image is generated by adapting the content from one image and the style from another, it is unusual to generate an image that completely matches both criteria [30]. Therefore, the use of weights ($\lambda_C$, $\lambda_S$, and $\lambda_A$) becomes very important to achieve a balance between the required style and content reconstruction. According to our observation, the higher the value of $\lambda_S$ the closer the style of the generated image to the reference image, while the higher $\lambda_C$ the closer the content of the generated image to the input image. Due to the fact that adversarial loss weight, $\lambda_A$, looks at an image from a higher level and compares how similar the synthetic images are to the field-taken images, we gave the most weight to adversarial loss weight. Fig. 7 presents some synthetic images generated with different combinations of $\lambda$ weights.

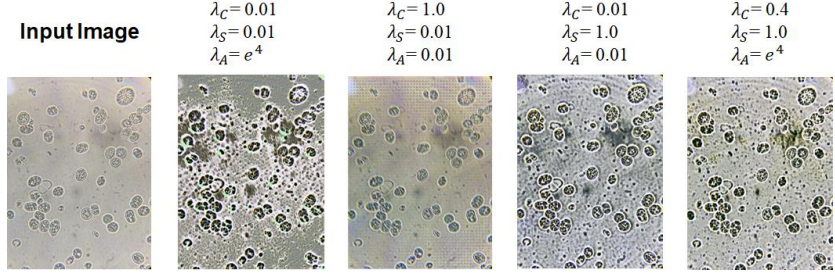

**Fig. 7**. Synthetically generate images with different $\lambda$ value combinations.

BioGAN was trained with 20 laboratory and 20 field images of *Prototheca bovis* for 100 epochs. Due to the memory constraint, all input and output images were set to the fixed size of $1024 \times 768$ for training. Fig. 8 shows samples of synthetic field images generated by BioGAN and two other baselines (see Results section for more details). The training time is highly dependent on the image size and the generator architecture. In the case of our proposed generator architecture, each iteration of each epoch took around 192 seconds for the U-Net (stride of 1 on convolutional layers), and 13 seconds for the Res-Net generator on a CUDA enabled NVIDIA GT730 GPU. For the testing, the inference time of BioGAN was found to be ~92 seconds.

## 3. Results

In this section we introduce firstly (subsection 3.1) the images we have used to train and evaluate the performance of our model. Specifically, to evaluate the fitness of the synthetic images produced by our model, we compare the results of our BioGAN algorithm with two other baselines, which have been used for unpaired image translation [7], and for transferring images' styles [32]. Given that the first baseline (CycleGAN) uses Cycle-Consistency loss (i.e. *Adversarial* loss in conjunction with pixel-level loss), and that the second baseline (Fast-style-transfer) uses *Perceptual* loss, while our model uses both, the comparison of the three models can help us to discriminate the contribution of each loss in generating microbiology field-like images. We present the results of a qualitative comparison in subsection 3.2.1, and the results of a qualitative comparison in section 3.2.2.

*3.1. Data collection and preparation*

We collected a dataset of bright-field microscopic laboratory images of *Prototheca bovis*, as well as field images of the same parasite. As it can be seen in Fig. 8, Prototheca's visual characteristics change significantly when grown in the laboratory or in the field, thus introducing additional challenges for I2IT. Laboratory samples are clean parasites that were grown in a laboratory environment, while field samples were produced by growing parasites in pig stool. The process of data collection was run and supervised by experienced biologists. In this study, 40 laboratory-taken images and 40 field images of *Prototheca bovis* were captured with a VWR IT 404 Inverted microscope's ocular lens (optical magnification of 400X and resolution of 4032 H×3024V).

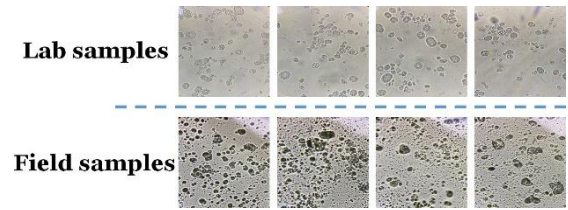

**Fig. 8.** Example images of *Prototheca bovis* parasites. Top row: laboratory samples. Bottom row: field samples

In total, 80 images were collected, 40 laboratory images with 1358 parasites, and 40 field images with 2899 parasites. These images were then annotated in COCO format [42] that enables us to train the object detection algorithms as explained in the following subsection, 3.2.

*3.2. Performance evaluation*

Following the training step explained in section 2.3 for our model, we fed 40 laboratory-taken images to the model for generating synthetic images. Similar procedures were applied for the two baseline models to generate synthetic images. We used both qualitative and quantitative approaches to evaluate the synthetic images generated by our model and by the two baselines.

Qualitative evaluations involve asking human raters to assess the realism of the generated images, [6,7], as explained in subsection 3.2.1. Because of the nature of our unpaired I2IT model and the absence of reference images, we had to quantify the quality of the synthetic images via a supervised object detection algorithm as explained in subsection 3.2.2.

*3.2.1 Qualitative evaluation*

Fig. 9 shows three original laboratory images and the corresponding synthetic images (generated by our model, CycleGAN, and Fast Style Transfer), which are supposed to look like target images (field images, one reported in Fig. 9 for comparison).

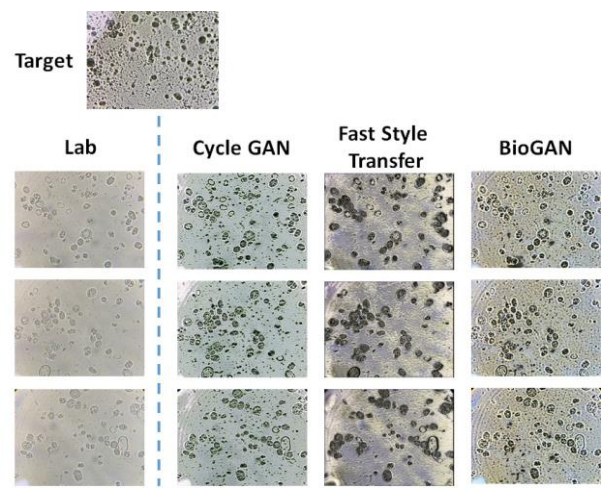

**Fig. 9.** Example of laboratory-taken images with their corresponding translation under the three models a) CycleGAN b) Fast-Style-Transfer c) BioGAN

Fig. 10 reports close-ups of synthetic images from the three models. Figs. 9 and 10 reveal the difference in the background structures of the synthetic images generated by BioGAN, Fast Style Transfer, and CycleGAN, respectively: the images generated via the Fast Style Transfer model seem to have a smoother structure with less scattered debris when compared with images from the other two models. Still, visual inspection shows that the background of CycleGAN and BioGAN synthetic images is more similar to the background of target images. Furthermore, the contents' contrast/gamma of Fast Style Transfer and BioGAN images seem to be more similar to the target images, when compared with CycleGAN images; this is arguably due to the application of *Perceptual* loss.

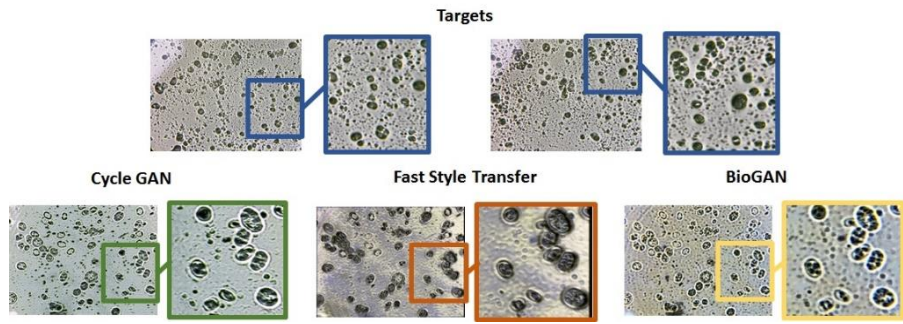

**Fig. 10**. A laboratory-taken image translated to a field-like image via three models.

Some studies, such as [7], used AMT (Amazon Mechanical Turk) for qualitative evaluation, with a public participant pool rating the synthetic images. For our study, using a public participant pool would not have been achievable because of the need for domain knowledge, due to the nature of parasite images; this motivated the use of two experienced biologists for our qualitative evaluation, reported Table 2.

**Table 2**. Qualitative evaluation of synthetic images from the three models of BioGAN, Fast Style Transfer, and CycleGAN; ratings by two expert biologists, from zero to ten, where zero means lowest similarity between synthetic and target image, and ten means highest similarity. Means and standard deviations based on ratings of 40 synthetic images.

|  | CycleGAN | Fast Style Transfer | BioGAN |
|---|---|---|---|
| Mean score | 7.2 $\pm$ 1.9 | 7.6 $\pm$ 1.6 | 8.3 $\pm$ 1.3 |

Biologists were shown 40 groups of 3 randomly sorted synthetic images, each image generated by one of the three models. Biologists (raters) were asked to rate the similarity between each image and a target image from zero to ten, where zero represents the lowest similarity and ten represents the highest similarity. Table 2 reports the means and standard deviations of the ratings for each model. Table 2 seems to suggest that Fast Style Transfer and BioGAN score better than CycleGAN. To evaluate whether the biologists rates (on the images that generated by BioGAN, CycleGAN, and Fast Style Transfer) are significantly different, a Wilcoxon test was used for the following combinations; i.e. BioGAN vs CycleGAN ($p < 0.001$), and BioGAN vs Fast Style Transfer ($p < 0.001$). This indicates that the synthetic images generated by BioGAN are rated significantly higher than those generated by the two other baselines.

Finally, to evaluate the association between the biologists' rates, we applied the Intraclass Correlation (ICC) test, using two-way mixed effect, consistency, and multiple raters' mode [45]. This test yielded a ICC3k value of 0.63 ($p=0.001$) for the rates of BioGAN images, which shows a moderate to good agreement between the biologists' rates for the quality of synthetic images generated by BioGAN.

*3.2.2 Quantitative evaluation*

To quantify the similarity of the synthetic images generated by *BioGAN* to the target images as compared to the synthetic images generated by the baselines, we used a MRCNN object detection framework, [43]. Specifically, we trained four MRCNN frameworks separately on laboratory images and three groups of synthetic images generated by the three models, and we tested the four MRCNN frameworks on parasite detection tasks with field images. The four MRCNN frameworks were trained with no augmentation and under the same conditions, i.e. with the same hyper-parameters and tested on the 40 field images (see section 3.1). The four frameworks were trained with two epochs including 500 iterations. For the object detection task *Precision*, *Recall*, *F1-score* and *mAP* per image were used to verify the performance of the three frameworks. They are defined as follows:

$$Precision = \frac{Tp}{Tp+Fp} \quad (6)$$

and

$$Recall = \frac{Tp}{Tp+Fn} \quad (7)$$

where *Tp* is the number of correctly detected (true positive) objects, *Fp* is the number of wrongly detected (false positive) objects, and *Fn* is the number of missed objects through the detection process (false negative). The higher the *Precision*, the more confident the model is about its detection, and the higher the *Recall*, the more objects the model has correctly detected. Due to the inherent trade-off between *Precision* and *Recall*, we calculated the *F1-score* which is a metric that measures the balance of *Precision* and *Recall*:

$$F1-score = \frac{Tp}{Tp+1/2\ (Fp+Fn)} \quad (8)$$

The *mAP* (Mean Average precision) is another important evaluation metric that has been widely used in world-class object detection challenges, including Pascal VOC [44] or COCO [42] to evaluate the performance of object detection models. We compute the VOC-style *mAP*, which computes the *Precision* and *Recall* for different detection thresholds [41] and measures the area under the Precision-Recall curve for each class (in our case there is only one class) at a certain IoU level (@IoU=70). As shown in Table. 3, the *mAP* for BioGAN is better than that of two other basslines which shows the better performance of the BioGAN-trained object detection framework at different level of detection confidence score. Table 3 also shows, *Precision*, *Recall*, *F1-score* and *mAP* of the four MRCNN frameworks trained separately on the laboratory images and the synthetic images generated by the three models (a constant *detection_minimum_confidence* parameter of 70% was used; any detection with the confidence score above 70% would be considered positive).

**Table 3.** Quantitative evaluation of the four MRCNN frameworks trained separately on the laboratory-taken images and synthetic images generated by the three models.

| Metric | Laboratory images (%) | CycleGAN synthetic images (%) | FST synthetic images (%) | BioGAN synthetic images (%) |
|---|---|---|---|---|
| Precision | 78.2 | 76.3 | **79.9** | 69.9 |
| Recall | 10.3 | 9.9 | 14.4 | **17.6** |
| F1-Score | 18.2 | 17.5 | 24.4 | **28.1** |
| $mAP \times 100$ | 8.1 | 7.6 | 11.5 | **12.3** |

Due to many undetected parasites, the *Recall* values for any framework are low. This is because of the significant difference in morphology between parasites grown in the laboratory and in the field. However, the use of BioGAN synthetic images results in the highest *Recall* value (17.6%). On the other hand, the *Precision* value, which represents the number of truly detected parasites, is lowest for BioGAN. *F1-scores* and *mAPs* are better metrics for the model's performance as they measure the balance between *Precision* and *Recall*. The *BioGAN*-trained framework shows an improvement, as compared to the laboratory trained framework, of +54.4% for *F1-score*, and +51.8% for *mAP*, respectively. Fast Style Transfer trained framework shows improvements of +34% (F1-score) and +41.9% (*mAP*), while CycleGAN trained framework shows lower values of *F1-score* and *mAP*. A significance analysis of *F1-scores* from BioGAN with the Fast Style Transfer, and CycleGAN was performed using a paired-sample Wilcoxon test. These tests yielded values of *p<0.002*, which demonstrate the significance of the BioGAN *F1-scores* in comparison with that of two other baselines.

We also tested whether training the frameworks with laboratory-taken images and synthetic images could increase the *Precision*, and consequently *F1-score* and *mAP*. For this purpose, three MRCNN frameworks were re-trained with laboratory and synthetic images. The evaluation results are reported in Table 4.

**Table 4.** Quantitative evaluation of the object detection frameworks, trained on a batch of laboratory and synthetic data.

| Metric | Laboratory images (%) | CycleGAN synthetic + laboratory images (%) | FST synthetic + laboratory images (%) | BioGAN synthetic + laboratory images (%) |
|---|---|---|---|---|
| Precision | 78.2 | **80.8** | 73 | 72.6 |
| Recall | 10.3 | 10.6 | 16.2 | **19.4** |
| F1-Score | 18.2 | 18.7 | 26.5 | **30.6** |
| $mAP \times 100$ | 8.1 | 8.6 | 11.9 | **14.2** |

The frameworks trained with CycleGAN+ laboratory data shows an improvement in *Precision*. The BioGAN+ laboratory trained framework has achieved a relative improvement of 68.1% (*F1-score*) and 75.3%

(*mAP*) as compared to laboratory trained framework. The significance of the *F1-score* for BioGAN+ laboratory-trained framework when compared with the other baselines (CycleGAN+ and Fast Style Transfer+ laboratory-trained frameworks) was examined using Wilcoxon tests, with the results indicating the significance of the BioGAN+ laboratory-trained framework *F1-score* ($p<0.002$).

## 4. Discussion

The collection of field images of parasites, like other biological systems, can be labour-intensive and costly. Although standard translation architectures like [7,19,32] have been implemented in some literature for staining or translating microscopy images [25,30,31,45,46], in this study, we developed for the first time a new GAN-based model (called BioGAN) to translate laboratory-taken microbiological images into field-like images. Due to the nature of microscopic image translation, which is an unpaired image translation problem, we utilized a Perceptual loss in conjunction with the *Adversarial* loss, to compensate for the absence of pixel-level loss in unpaired data problems. This paper analyses how *Adversarial* and *Perceptual* loss contribute to the generation of realistic-looking synthetic images by comparing our BioGAN model with CycleGAN, a model that utilizes *Adversarial* and Cycle-Consistency loss that forms the basis of some other image translation studies [27,28,30,31], and with Fast Style Transfer [32], a model that employs *Perceptual* loss alone. We have shown that *Perceptual* loss is able to transfer a fixed-style texture through the entire image, which helps translate the background of the laboratory images into field-like images. We have also shown that the *Adversarial* loss can encourage the generator in the GAN network to create a more realistic cell morphology, which is found in field images (see Fig. 9). The proposed BioGAN model has shown the ability to transfer from the laboratory images meaningful spatial features, such as object's boundaries, along with meaningful style features (i.e. texture) from the field images.

Quantitative evaluation has shown that an object detection framework trained on the synthetic images generated by BioGAN results in a slight reduction in *Precision* and in an improved *Recall*, as compared to a framework trained on laboratory images only. An increase in the *Recall* means there are fewer parasite cells missed by the object detector; this can be viewed as evidence that by BioGAN synthetic images are more similar to field images when compared with [19,32]. However, a lower Precision means that the framework is detecting spurious objects as parasites. BioGAN's synthetic images, also, resulted in an improvement of 54.3% and 51.8%, for *F1-Score* and *mAP*, respectively, when compared to a framework trained on laboratory images only. This improvement increases when the framework is trained on BioGAN synthetic images and laboratory images simultaneously.

*4.1. Limitations*

This work shows that BioGAN is able to generate synthetic images which are similar to the field images, but important challenges remain. Real field microscopic images often contain random objects (i.e. unprocessed foods in stool samples, or contamination in water samples) and debris, which currently cannot be synthesized by our image translation model. Our observations indicate that the presence of random objects deteriorates the performance of the object detection frameworks, because they can cause false positive detection in some cases. In order to have a model that is able to transfer these random objects we might require a more content-aware functionalities that can intelligently generate and harmonize the random objects into the synthetic image. We believe that this can be an interesting topic of research which can open the application of image translators to other biological systems. Another limitation of the current work is the high runtime, which is due to the complexity of the network. We acknowledge that the real-time inference for microbiological images may not be a necessity, but in the case that practitioners wanted to apply BioGAN on other domains, that may require a faster execution. Therefore, it may be important to explore the possibility of further optimizing the network in future.

## 5. Conclusion

In this paper, we present an unpaired, unsupervised model, namely BioGAN, for translation of microbiological images, inspired by Cycle-GAN [7] and Fast Style Transfer [32]. By utilizing *Adversarial* and *Perceptual* loss, the CNN-based generator was able to transfer the visual characteristics of a reference (i.e. field-taken) image to the laboratory-taken image. The proposed model was tested on its ability to translate laboratory-taken images of *Prototheca bovis* into field-like images, using experts' qualitative evaluation and quantitative evaluation by the MRCNN object detection framework. When compared to two other baselines, both qualitative and quantitative evaluations of the model show promising results in generating realistic field-like images. We have now made the BioGAN network publicly available on our GitHub repository, and our future efforts will be directed towards investigating the generalizability of BioGAN to translate other medical imaging modalities.

**Acknowledgements**

This research did not receive any specific grant from funding agencies in the public, commercial, or not-for-profit sectors